\documentstyle[12pt]{article}

\newcommand{\aaa}{{\cal A}}

\newcommand{\be}[1]{\begin{equation}\label{#1}}
\newcommand{\bra}[1]{{\langle #1 |}}
\newcommand{\braket}[2]{{\langle #1 |\, #2\rangle}}

\newcommand{\ee}{\end{equation}}

\newcommand{\ket}[1]{{|\, #1\rangle}}
\newcommand{\ketbra}[2]{{|\, #1\rangle\langle #2|}}

\newcommand{\omg}{\Omega}
\newcommand{\rrr}{{\cal R}}

\newcommand{\spn}{{\rm span}}

\title{An Algebraic Quantization of Causal Sets}

\author{Ioannis Raptis\footnote{Algebra and Geometry Section,
    Department of Mathematics, University of Athens,
    Panepistimioupolis 157 84, Greece, e-mail:
    iraptis@eudoxos.dm.uoa.gr}}
    
\date{}

\begin{document}

\maketitle

\begin{abstract}
A scheme for an algebraic quantization of the causal sets of Sorkin {\it et al.} is presented. 
The suggested scenario is along the lines of a similar algebraization and quantum interpretation of 
finitary 
topological spaces due 
to Zapatrin and this author. To be able to apply the latter procedure to causal sets Sorkin's 
`semantic switch' from `partially ordered sets as finitary topological spaces' to `partially ordered 
sets as locally finite causal 
sets' is employed. The result is the definition of `quantum causal sets'. Such a procedure  
and its resulting definition is physically justified by a property of quantum causal 
sets that meets Finkelstein's 
requirement from `quantum causality' to be an immediate, as well as an algebraically represented, 
relation 
between events for discrete 
locality's sake. The quantum causal sets introduced here are shown to have this property by direct use 
of a  
result from the algebraization of finitary topological spaces due to Breslav, Parfionov and Zapatrin.  
\end{abstract}
\vskip 0.1in
\noindent {\bf 1. INTRODUCTION}
\vskip 0.1in
An effective procedure has been developed for substituting a continuous topological space, 
such as a bounded region in a spacetime manifold, by a finitary one which was then seen to possess 
the structure of a partially ordered set (poset) (Sorkin, 1991). 
With every such poset an algebra, the poset's incidence algebra, was subsequently associated 
(Breslav, Parfionov and Zapatrin, 1999). Hence finitary substitutes for continuous topologies enjoyed a 
purely algebraic 
representation 
in terms of incidence algebras. Recently a quantum interpretation was given to 
the latter algebraized finitary topological spaces and the whole procedure was called `algebraic 
quantization 
of discretized spacetimes' (Raptis and Zapatrin, 1999).

On the other hand Sorkin has accounted for a significant change of physical interpretation for the 
aforementioned 
posets, from ones whose partial order encodes topological information as in (Sorkin, 1991), to ones 
whose partial 
order stands for the causal, `after' relation between events (Sorkin, 1989). Thus he and co-workers 
arrived at the 
notion of causal set (Bombelli {\it et al.}, 1987 and Sorkin, 1989).  
The new interpretation also substituted the posets' finitarity by the causal sets' local finiteness 
property. 
Thus it seems inevitable or at least natural that the incidence algebras associated with the 
topologically interpreted posets inherit this causal interpretation from the `semantic switch' 
advocated by Sorkin (1989). The resulting algebras may be coined `causal incidence algebras'. 
Finally, if we give these algebras a quantum interpretation as in (Raptis and Zapatrin, 1999), 
we are naturally led to `quantum causal incidence algebras'. Then effectively we 
will have algebraically quantized Sorkin {\it et al.}'s causal sets to `quantum causal sets'.

These quantum causal incidence algebras are presented here as sound models 
of quantum causal sets. We support our claim by a result from (Breslav, Parfionov and Zapatrin, 1999) 
that vindicates 
an insight of Finkelstein (1988) on how to model causality in the quantum deep.

The paper is organized as follows: in section 2 we recall some key results from (Sorkin, 1991), 
essentially how a 
finitary substitute for continuous topology has the structure of a poset. In section 3 we recall from 
(Breslav, Parfionov and Zapatrin, 1999) how 
to associate with every poset an algebra-the poset's incidence algebra. In section 4 we briefly present 
Sorkin's 
semantic switch from `posets as finitary topological spaces' to `posets as locally finite causal sets' 
found in (Sorkin, 1989), 
thus define 
causal incidence algebras. In section 5 we select from (Raptis and Zapatrin, 1999) some aspects of the 
quantum 
interpretation of incidence 
algebras, hence lead to the structure of quantum causal incidence algebras modelling quantum causal 
sets. In the last section 6  
we use a result from (Breslav, Parfionov and Zapatrin, 1999) that supports the soundness of our 
algebraic models of quantum causal sets.  
The conclusion at the end resumes our approach to quantum causal sets.
\vfill\eject
\noindent {\bf 2. ASPECTS OF FINITARY SUBSTITUTES}
\vskip 0.1in
The essential result from (Sorkin, 1991) for our exposition here is the equivalence between finitary 
substitutes of 
bounded regions of continuous topological spaces and posets. Below we recall briefly this equivalence. 

Assume a finite continuous topological space $S$, for instance, a bounded region of a spacetime 
manifold. Let $S$ 
be covered by a locally finite collection ${\cal{U}}$ of bounded open sets $U$ in the sense that each  
of $S$'s points has an open neighborhood that meets only a finite number of $U$s in ${\cal{U}}$. 
Any two points $x,y$ of $S$ 
are indistinguishable with respect to its locally finite open cover ${\cal{U}}$ if $\forall U\in 
{\cal{U}}:\, x\in U\Leftrightarrow y\in U$. Indistinguishability with respect to this subtopology 
${\cal{U}}$ 
of $S$ is an equivalence relation on the latter's points and is symbolized by $\leftrightarrow$. Taking  
the quotient $S/\!{\leftrightarrow}$ results in the substitution of $S$ by equivalence classes 
of its points whereby two points in the same equivalence class are covered by (ie belong to) the same, 
finite in number, open neighborhoods $U$ of ${\cal{U}}$, thus are indistinguishable by it. Call 
the quotient space ${\cal{F}}$.

Now let $x,y$ be points belonging to two distinct equivalence classes in ${\cal{F}}$. Consider the 
smallest 
open sets in the subtopology ${\cal{U}}$ of $S$ containing $x$ and $y$ respectively given by: 
$\Lambda(x):=\cap\{ U\in {\cal{U}}:\, 
x\in U\}$ and $\Lambda(y):=\cap\{ U\in {\cal{U}}:\, y\in U\}$. Define the relation $\rightarrow$ 
between 
$x$ and $y$ as follows: $x\rightarrow y\Leftrightarrow \Lambda(x)\subset\Lambda(y)\Leftrightarrow 
x\in\Lambda(y)$. 
Then assume that $x\leftrightarrow y$ in the previous paragraph stands for $x\rightarrow y$ and 
$y\rightarrow x$.   
$\rightarrow$ is a partial order on ${\cal{F}}$ and $S$ has been effectively substituted by the 
finitary ${\cal{F}}$ which is a $T_{0}$ topological space with the structure of a poset. Sorkin uses the 
finitary topological and partial order theoretic languages interchangeably exactly due to this 
equivalence between $T_{0}$ finitary substitutes and posets. For the future purposes of the present 
paper we distill 
this to the following statement: in (Sorkin, 1991) a partial order is interpreted topologically. We call it 
`topological partial 
order' and the poset encoding it `topological poset'.
\vskip 0.1in
\noindent {\bf 3. ASPECTS OF INCIDENCE ALGEBRAS}
\vskip 0.1in
The aspect of (Breslav, Parfionov and Zapatrin, 1999) that is of significance here is that with every 
topological poset 
$P$ an algebra $\omg(P)$-the poset's incidence algebra-is
associated, so that the order theoretic encodement of finitary substitutes has an equivalent algebraic 
description in terms of incidence algebras. $\omg(P)$ as a linear space, in Dirac's ket-bra 
notation\footnote{The 
reader can also refer to the paper of 
Zapatrin (1998) for an early and detailed exposition of incidence algebras associated with topological 
posets. However, 
incidence algebras were not presented there in Dirac's ket-bra notation, so in this paper we solely 
refer to (Breslav, Parfionov and Zapatrin, 1999) that first exposed them in such a way.}, 
is defined as

\be{e11}
\omg(P)=\spn\{\ketbra{p}{q}:\, p\rightarrow q\} , 
\ee

\noindent with product between two of its ket-bras given by

\be{e12}
\ketbra{p}{q} \cdot \ketbra{r}{s} =
\ket{p} \braket{q}{r} \bra{s} =
\braket{q}{r} \cdot \ketbra{p}{s} =
\left\lbrace \begin{array}{rcl}
\ketbra{p}{s} &,& \mbox{if } q=r \cr
0 && \mbox{otherwise.}
\end{array} \right.
\ee

In a so-called `spatialization procedure' Breslav, Parfionov and Zapatrin brought into 1-1 
corespondence the elements $p$ of a 
poset $P$ and the primitive ideals $I_{p}$ of its incidence algebra $\omg(P)$ by defining the latter as

\be{e13}
I_{p}=\{\ketbra{q}{r} :\, \ketbra{q}{r}\not= \ketbra{p}{p}\} ,
\ee

\noindent thus defined the primitive spectrum of $\omg(P)$ as ${\cal{S}}=\{ I_{p}\}$.

Then the Rota topology is defined on ${\cal{S}}$ as being generated by the following relation $\rho$ 
between 
any two primitive ideals $I_{p},I_{q}\in{\cal{S}}$  

\be{e14}
I_{p}\rho I_{q}\Leftrightarrow I_{p}I_{q}(\not= I_{q}I_{p})\subset I_{p}\cap I_{q} ,
\ee

\noindent with $I_{p}I_{q}$ their product ideal.

The central question being raised and settled in (Breslav, Parfionov and Zapatrin, 1999) is when the 
Rota topology on 
${\cal{S}}$ of $\omg(P)$ is the same as  
the finitary one when the poset $P$, whose incidence algebra is $\omg(P)$, is the finitary substitute of 
a continuous 
topological space, a topological poset, as in section 2. The answer to this question provided an 
invaluable clue which is used 
in section 6 for showing that 
a quantum causal incidence algebra has an important property, deriving from considerations of discrete 
locality, 
that any sound 
algebraic model of `quantum causality' should possess. Thus we postpone its presentation until section 
6. 
First we need to make contact with causality by following Sorkin's paradigm of reinterpreting posets 
from topological  
to causal.
\vfill\eject
\noindent {\bf 4. SORKIN'S CAUSALIZATION OF POSETS}
\vskip 0.1in
In a change of physical interpretation, ultimately of physical theory, Sorkin stopped thinking of posets 
as 
encoding the topological 
information of finitary substitutes of continuous topological spaces and reinterpeted the partial order 
$\rightarrow$ 
between their elements as the causal, `after' relation between events. In a revealing paper (Sorkin, 
1989) 
he recalled this 
semantic switch of his as follows: \begin{quotation}...Still, the order inhering in the finite topological 
space seemed to be very 
different from the so-called causal order defining past and future. It had only a topological meaning but 
not 
(directly anyway) a causal one.
In fact the big problem with the finite topological space was that it seemed to 
lack the information which would allow it to give rise to the continuum in all its aspects, not just in the 
topological aspect, but with its metrical (and therefore its causal) properties as well...The way out of 
the 
impasse involved a conceptual jump in which the formal mathematical structure remained constant, but 
its physical 
interpretation changed from a topological to a causal one...The essential realization then was that, 
although order 
interpreted as topology seemed to lack the metric information needed to describe gravity, the very same 
order 
reinterpreted as a causal relationship, did possess information in a quite straightforward sense...In fact 
it took me 
several years to give up the idea of order-as-topology and adopt the causal set alternative as the one I 
had been 
searching for...\end{quotation} 

This significant change of the physical semantics of the same mathematical structure, the poset, 
amounted to  
the latter being interpreted by Sorkin and co-workers as a causal set: `{\it a locally finite set of points 
endowed with a partial order corresponding to the macroscopic relation that defines past and future}'  
(Bombelli {\it et al.}, 1987 and Sorkin, 1989). Local finiteness was defined as 
follows: use $\rightarrow$ of a poset $P$, interpreted now as a causal relation on the causal set, 
to redefine $\Lambda(x)$ 
of section 2 for some $x\in P$ as $\Lambda(x)=\{ y\in P:\, y\rightarrow x\}$,  
and dually $V(x)=\{ y\in P:\, x\rightarrow y\}$. $\Lambda(x)$ is the `causal past' of the event $x$ and 
$V(x)$ its 
`causal future'. Local finiteness then requires the so-called Alexandroff set $V(x)\cap \Lambda(y)$ to 
be finite for all 
$x,y\in P$ such that 
$x\in \Lambda(y)$. In other words, only a finite number of events `causally mediate' between any two 
events 
$x,y$, with $x\rightarrow y$, of the causal set $P$. Roughly, the finitarity of the topological posets of 
section 2 translates 
by Sorkin's semantic switch to the local finiteness of causal sets, although it must be stressed that the 
physical theories that 
they support, the topological 
discretization of manifolds in (Sorkin, 1991) and causal set theory in (Bombelli {\it et al.}, 1987 and 
Sorkin, 1989) respectively,  
are quite different in motivation, scope and aim (Sorkin, 1989 and 1991). 

Here we follow Sorkin's example and advocate a similar semantic switch from `incidence algebras 
$\Omega(P)$ 
associated with topological posets $P$', to `incidence algebras $\omg(P)$ associated with causal sets 
$P$'. That is 
to say, we change physical meaning for the arrow $p\rightarrow q$ encoded in the ket-bra notation in 
$(1)$ from 
topological  
to causal. Thus the proposed change of meaning is from `topological incidence algebras' to `causal 
incidence algebras'. 
However, in the new algebraic environment of incidence algebras, apart from the $\rightarrow$ 
structure of $P$ which is 
effectively encoded into the ket-bra symbol and the product of ket-bras in $\omg(P)$ $(2)$, a new 
element of structure absent 
from $P$, namely superposition $+$ of ket-bras in $\Omega(P)$, enables us to also impart quantum 
interpretation to 
$\omg(P)$ no matter whether the latter is of topological or of causal nature. We present elements of 
this theory next.
\vskip 0.1in
\noindent {\bf 5. QUANTIZATION OF INCIDENCE ALGEBRAS}
\vskip 0.1in
In (Raptis and Zapatrin, 1999) a quantum interpretation to topological incidence algebras was given, 
thus the 
authors arrived 
at an algebraically quantized version of Sorkin's (1991) discretized spacetimes. The reader is referred 
to (Raptis and Zapatrin, 1999) for technical details. Below we only collect from it the evidence 
supporting this 
quantum interpretation of incidence algebras:

a. The algebraic operation $+$ between ket-bras in $\omg(P)$ naturally enjoys a physical interpretation 
as coherent quantum superposition.

b. The split of any incidence algebra $\omg(P)$ into a commutative subalgebra ${\cal{A}}$ of grade 
zero 
vectors and a linear subspace ${\cal{R}}$ of vectors with grade $\geq1$ as $\omg = \aaa \oplus \rrr$, 
naturally affords a physical interpretation as the algebra of quantum spacetime states (called 
stationaries in that paper) and 
the space of quantum dynamical transition processes between them (called transients and paths in that 
paper), respectively.

c. The quantum interpretations given in a and b above seem all the more plausible when one also 
interprets Sorkin's 
`inverse limit' of finitary substitutes $P$ to the continuous manifold space $M$ that they approximate 
as a correspondence limit 
in the quantum sense of Bohr. Then ${\cal{A}}$ is expected to decohere to the classical commutative 
algebra of spacetime 
coordinates parametrizing events (classical `position' vector states), while ${\cal{R}}$ to the classical 
cotangent 
Lie algebra of kinematical derivations of them (classical `momentum' covector states).

What is of importance here is to borrow the quantum interpretation of topological incidence algebras 
from 
(Raptis and Zapatrin, 1999) and 
apply it to the causal incidence algebras of the previous section. Doing so we arrive straightforwardly 
at the 
concept of `quantum causal incidence algebras' modelling `quantum causal sets'. The soundness of this 
model of 
quantum causal sets is shown next.
\vskip 0.1in
\noindent {\bf 6. LOCAL ALGEBRAIC QUANTUM CAUSALITY}
\vskip 0.1in
Finkelstein (1988) intuited that a sound quantum model of causality should essentially 
meet the following two conditions:

a. Be algebraic and have a quantum interpretation for this algebraic structure, and

b. What is algebraized should not be the classical causality relation which, like the one between the 
elements of Sorkin's 
causal sets, is usually modelled by a partial order which in turn, being transitive (Bombelli {\it et al.}, 
1987 and Sorkin, 1989), 
is non-local (mediated). Rather, a local 
(immediate) version of it should be algebraically quantized. That is, the physical causal connection 
between events in the 
quantum deep should be one connecting 
nearest-neighbouring events. Symbolically, $\rightarrow$ is such that $(x\rightarrow y)$ and 
$\not\!\exists z:\, x\rightarrow z\rightarrow y$. 

Causal net theory was proposed as a local, discrete, algebraic and quantum interpreted alternative to 
Sorkin's causal set theory 
that satisfies the two demands above (Finkelstein, 1988). The discrete locality aspect of causal net 
theory is that it can be 
thought of as causal set theory constrained 
to Alexandroff neighborhoods $V(x)\cap\Lambda(y)$ ($x\rightarrow y$) of zero cardinality. Its 
quantum algebraic aspect is 
that it represents causal relations between events algebraically with a quantum interpretation for this 
algebraic 
structure. In brief, causal nets are sound models of `quantum causal spaces'. In the 
present paper too 
quantum causal incidence algebras are proposed as models of quantum causal sets and they plainly 
meet Finkelstein's 
requirement a. Below we show how they also satisfy b in a straightforward way.

Recall the question posed at the end of section 3. Breslav, Parfionov and Zapatrin (1999) showed that 
the topological 
information encoded 
in a topological poset $P$ due to Sorkin (1991) is the same as that encoded in the Rota topology $\rho$ 
$(4)$ on the primitive 
spectrum ${\cal{S}}$ of its associated incidence algebra $\omg(P)$ if and only if the elements $p,q$ 
of the poset, 
indexing as in $(3)$ the primitive ideals $I_{p}$ and $I_{q}$ related by $\rho$ in $(4)$, 
are immediately connected by $\rightarrow$ 
in the Hasse diagram of $P$; that is to say, if and only if $(p\rightarrow q)$ and 
$\not\!\exists r:\, p\rightarrow r\rightarrow q$. As we have causally interpreted 
$\rightarrow$ in $P$ \`{a}-la Sorkin, this theorem shows that causal incidence algebras 
are sound models of local 
causal sets and, {\it in extenso}, quantum causal incidence algebras of quantum causal sets, according 
to Finkelstein's two 
basic requirements presented above.
\vskip 0.1in
\noindent {\bf 7. CONCLUSION}
\vskip 0.1in
We may resume the algebraic quantization procedure leading to quantum causal incidence algebras in 
the following diagram

\[
\begin{array}{rcl}
&{\rm t-posets/incidence\,\, algebras}\,{\buildrel{q}\over{\longrightarrow}}\,{\rm q-t-
posets/incidence\,\, algebras}\cr
&c\downarrow\hskip 2.6in \downarrow c\cr
&{\rm c-posets/incidence\,\, algebras}\,{\buildrel{q}\over{\longrightarrow}}\,{\rm q-c-
posets/incidence\,\, algebras}\cr
\end{array}
\]

\noindent picturing from the upper left corner the processes of `(c)ausalization' (causal re-interpretation 
\`{a}-la Sorkin) 
of (t)opological posets (t-posets) and their topological incidence algebras to causal sets (c-posets) and 
their 
causal incidence algebras, 
followed by `(q)uantization' (quantum interpretation according to the Raptis-Zapatrin scheme) to 
quantum causal incidence algebras modelling quantum causal sets (q-c-posets). One may equivalently 
follow the other route and 
first quantize topological posets 
and their topological incidence algebras to quantum topological posets (q-t-posets) and their quantum 
topological incidence algebras and 
then causalize them to quantum causal sets. In this paper we took the c-followed-by-q route.
\vskip 0.1in
\noindent {\bf ACKNOWLEDGMENTS}
\vskip 0.1in
Early exchanges on `quantum causality' with Professor Anastasios Mallios, as well as recent 
discussions on finitary substitutes 
and the algebraic quantization of their incidence algebras with Professor Roman Zapatrin, are kindly 
acknowledged.
\vfill\eject
\noindent {\bf REFERENCES}
\vskip 0.1in
\noindent Bombelli L., Lee J., Meyer D. and Sorkin R. D. (1987). Space-Time as a Causal Set, 
{\it Physical Review Letters}, {\bf 59}, 521.

\noindent Breslav R. B., Parfionov G. N. and Zapatrin R. R. (1999). Topology measurement within the 
histories approach, 
{\it Hadronic Journal}, {\bf 22}, 2, 225, e-print quant-ph/9903011.

\noindent Finkelstein D. (1988). `Superconducting' Causal Nets, {\it International Journal of 
Theoretical Physics}, 
{\bf 27}, 473.

\noindent Raptis I. and Zapatrin R. R. (1999). Quantization of discretized spacetimes and the 
correspondence principle, 
submitted to the {\it International Journal of Theoretical Physics}, e-print gr-qc/9904079.

\noindent Sorkin R. D. (1989). A Specimen of Theory Construction from Quantum Gravity in {\it The 
Creation 
of Ideas in Physics}, ed. Leplin J., Kluwer Academic Publishers, Dordrecht (1995).

\noindent Sorkin R. D. (1991). Finitary Substitute for Continuous Topology, {\it International Journal 
of Theoretical 
Physics}, {\bf 30}, 923.

\noindent Zapatrin R. R. (1998). Finitary Algebraic Superspace, {\it International Journal of 
Theoretical 
Physics}, {\bf 37}, 799, e-print qr-qc/9704062.

\end{document}